\documentclass[titlepage,12pt]{article}
\usepackage{epsfig,psfig}

\newcommand{\be}{\begin{equation}}
\newcommand{\ee}{\end{equation}}
\begin{document}
%
%
%\psdraft
%
%
%\begin{titlepage}
%\vspace*{4cm}
\begin{center}
{\bf \large SOLAR NEUTRINOS AS HIGHLIGHT OF ASTROPARTICLE PHYSICS}~
\footnote{invited lecture at 25th International Cosmic Ray Conference,
Durban, 28 July - 8 August, 1997}
\vskip 5mm
{\bf V. Berezinsky}\\
\vskip 5mm
INFN, Laboratori Nazionali del Gran Sasso, 67010 Assergi (AQ), Italy \\
\vskip 10mm
ABSTRACT
\end{center}
%\vskip 15mm
\noindent
Solar neutrinos are discussed in the light of the new data and of the recent 
progress in helioseismology. The most attention is given to the new status 
of Standard Solar Models due to seismically measured density and sound 
speed in the inner solar core. The elementary particle solutions to 
the Solar Neutrino Problem and their observational signatures are discussed
. \\*[3mm]
\noindent
1. INTRODUCTION\\*[3mm]
As everybody knows, the earth and sun were created for the sake of 
neutrino-oscillation experiment. The distance between them was chosen as 
vacuum oscillation length, the density inside the sun - to fit the MSW 
effect and the sun was prepared as an ideal pure $\nu_e$-source. When
all this was done, there were created Bruno Pontecorvo - to invent idea 
of neutrino oscillations, John Bahcall - to calculate the solar-neutrino flux,
Ray Davis - to perform the first solar-neutrino experiment and all other 
solar-neutrino people to accomplish this difficult job. The last in this 
row was Maury Goodman - to cite silly jokes like this one in his neutrino 
news.  

In this review I want to convince the readers, that the Solar Neutrino 
Problem (SNP) is not a vague astrophysical puzzle, but an 
elementary-physics problem - the first step beyond the Standard Model of 
EW interactions. The neutrino with the properties as in the Standard Model
will be referred to as {\em standard neutrino}.

These years we are entering the new epoch in solar-neutrino physics.\\
1. It is characterized by the new detectors: SuperKamiokande (in operation)
and SNO and Borexino, which will start to operate soon. These detectors 
will be able to discover the direct signatures of elementary particle 
solution to the SNP in a few years.\\
2. During last several years, the convincing model-independent
arguments demonstrated that there is no astrophysical solution (nuclear 
physics included) to the SNP. Now we go much further. Helioseismic data confirm
the Standard Solar Models (SSMs) with high precision at all radial 
distances at interest. The confidence to the SSMs is greatly 
strengthened and their serious revision looks now unrealistic. The 
interest is shifted from the model-independent analysis to the accuracy of 
the SSM 
predictions and especially to the fine details not essential for the 
prediction of neutrino fluxes.\\
3. The uncertainties in the cross-sections are now the dominant ones for 
the prediction of the solar-neutrino fluxes. The impressive progress 
exists here as well.  In the LUNA experiment at Gran Sasso the cross-section 
of one of the most intriguing reactions,$^3He+^3\!\!He \to ^4\!\!He+2p$, was 
measured at energy corresponding to maximum of the Gamow peak in the 
Sun. The famous speculations about solving or ameliorating the SNP due to 
increase of this cross-section at very low energy, have been now honorably 
buried.
In the nearest future most of cross-sections relevant to the SMP will be 
measured in the LUNA experiment at very low energies.\\*[1mm]

The plan of my review is as follows. I will start with the good 
old-fashioned story about  disfavouring (which is a good diplomatic equivalent 
for the word excluding) of astrophysical solution to the SNP.
In section 3 I will describe 
the status of the SSMs in the light of helioseismic data. In the section 
4 the status of nuclear-reaction cross-sections will be shortly reviewed.
In sections 5 and 6  the elementary-particle solutions to the SNP will be 
discussed.
For early detailed reviews see e.g. Bahcall 1989, Turck-Chieze et al 1993,
Bowles and Gavrin 1993, Kirsten 1995 and Castellani et al 1997.\\*[4mm]
\noindent 
2. STATUS OF ASTROPHYSICAL SOLUTION TO SNP\\*[4mm]
\noindent
The Solar Neutrino Problem (SNP) is a deficit of neutrino fluxes detected 
in all four solar-neutrino experiments (Suzuki 1997, Conner 1997, 
Hampel et al 1996, Abdurashitov et al 1997, Cleveland et al 1995).
The data, as reported up to 1997, 
are listed in Table 1 and compared with calculations of Bahcall and 
Pinsonneault (1995) for the Standard Solar Model (SSM).
\begin{table}[htb]  
\caption{
The solar-neutrino data of 1997 compared with the SSM prediction , Bahcall
and Pinnsoneault (1995)}.
\center{\begin{tabular}{||c|c|c|c||}
\hline 
                &  DATA              & SSM [5]             & DATA/SSM \\
\hline 
GALLEX\         &                    &                     &            \\
(SNU)           &$69.7\pm 6.7^{+3.9}_{-4.5}$ &137           &$0.509\pm0.059$\\
SAGE\           &                    &                     &            \\
(SNU)           &$69\pm10^{+5}_{-7}$  &137                 &$0.504\pm0.085$\\
SUPERK\     &                    &                     &             \\
($10^6~cm^{-2}s^{-1}$)&$2.44\pm0.06 ^{+0.25}_{-0.09}$ &6.5 &$0.368\pm0.026$\\
HOMESTAKE\      &                    &                     &              \\
(SNU)           &$2.55\pm0.17\pm0.18$ &9.3                  &$0.274\pm0.027$\\
\hline
\end{tabular}}
\end{table}
\par 
\vspace*{0.5cm}
The solar neutrino spectrum is characterized by three most important components
(see Table 2).

(i) The most energetic part of the spectrum is presented by boron neutrinos 
from\\ 
\mbox{$^8B \to ^8\!\!Be+e^++\nu_e$} decay. The maximum energy of neutrinos 
in this spectrum is $E_{\nu,max} \approx 14~MeV$. Kamiokande and 
SuperKamiokande detect only boron neutrinos.

(ii) Beryllium neutrinos ($^7Be+e^- \to ^7\!\!Li+\nu_e$) are monoenergetic
($E_{\nu}=0.862~MeV$). Predicted flux of Be-neutrinos depends weakly on 
the solar model and is $(4.2 - 4.9)\cdot 10^9~cm^{-2}s^{-1}$. Homestake 
detects both boron and beryllium neutrinos: $^8B$-neutrinos 
provide about $80\%$ of the total signal and $^7Be$-neutrinos - about 
$13\%$

(iii) The low-energy part of the solar-neutrino spectrum is presented by $pp$-
neutrinos ($p+p \to ^2\!\! H+e^++\nu_e$) with maximum energy 
$E_{\nu}^{max}=0.42~MeV$. These neutrinos greatly overnumber the flux of 
other neutrinos ($6.1\cdot10^{10}~cm^{-2}s^{-1}$). The flux is determined 
practically only by solar luminosity and therefore is reliably 
predicted. GALLEX and SAGE measure mostly pp-neutrinos (about $50\%$ of 
the total signal) with some contribution of $^7Be$ and $^8B$ neutrinos 
(about $28\%$ and $12\%$, respectively).

>From Table 1 one can see that the suppression factor, DATA/SSM, is 
less for Homestake than for SuperKamiokande detector.
It is easy to understand that it creates a problem for 
nuclear/astrophysical solution to the SNP. Indeed, since the nuclear/astrophysical
  factors cannot 
change the shape of $^8B$-neutrino spectrum, the suppression factor for 
boron neutrino signal in Homestake should be the same as in 
Kamiokande.
%\newpage
% GNUPLOT: LaTeX picture
\begin{table}[t]
\caption{Nuclear reactions in the Sun (pp-chain)}
\setlength{\unitlength}{0.240900pt}
\ifx\plotpoint\undefined\newsavebox{\plotpoint}\fi
\sbox{\plotpoint}{\rule[-0.200pt]{0.400pt}{0.400pt}}%
\begin{picture}(1500,1619)(0,0)
\font\gnuplot=cmr10 at 10pt
\gnuplot
\sbox{\plotpoint}{\rule[-0.200pt]{0.400pt}{0.400pt}}%
%\put(176.0,832.0){\rule[-0.200pt]{303.534pt}{0.400pt}}
%\put(806.0,68.0){\rule[-0.200pt]{0.400pt}{368.095pt}}
%\put(176.0,68.0){\rule[-0.200pt]{303.534pt}{0.400pt}}
%\put(1436.0,68.0){\rule[-0.200pt]{0.400pt}{368.095pt}}
%\put(176.0,1596.0){\rule[-0.200pt]{303.534pt}{0.400pt}}
\put(554,1520){\makebox(0,0)[l]{p + p}}
\put(838,1520){\makebox(0,0)[l]{D + e + $\nu_e$}}
\put(491,1290){\makebox(0,0)[l]{p + D}}
\put(809,1290){\makebox(0,0)[l]{$^3$He + $\gamma $}}
\put(365,1099){\makebox(0,0)[l]{I}}
\put(1121,1099){\makebox(0,0)[l]{II}}
\put(554,1176){\makebox(0,0)[l]{86 \%}}
\put(932,1176){\makebox(0,0)[l]{14 \%}}
%\put(239,985){\makebox(0,0)[l]{$^3$He + $^3$He }}
\put(139,985){\makebox(0,0)[l]{$^3$He + $^3$He }}
%\put(554,985){\makebox(0,0)[l]{$^4$He + 2 p}}
\put(454,985){\makebox(0,0)[l]{$^4$He + 2 p}}
%\put(932,985){\makebox(0,0)[l]{$^3$He + $^4$He}}
\put(832,985){\makebox(0,0)[l]{$^3$He + $^4$He}}
%\put(1184,985){\makebox(0,0)[l]{$^7$Be + $\gamma$}}
\put(1120,985){\makebox(0,0)[l]{$^7$Be + $\gamma$}}
\put(901,832){\makebox(0,0)[l]{14 \%}}
\put(1090,832){\makebox(0,0)[l]{0.02 \%}}
%\put(617,603){\makebox(0,0)[l]{$^7$Be +$ e^-$}}
\put(567,603){\makebox(0,0)[l]{$^7$Be +$ e^-$}}
%\put(869,603){\makebox(0,0)[l]{$^7$Li+$\nu_e$}}
\put(819,603){\makebox(0,0)[l]{$^7$Li+$\nu_e$}}
\put(1058,603){\makebox(0,0)[l]{$^7$Be + p }}
\put(1310,603){\makebox(0,0)[l]{$^8$B + $\gamma$}}
%\put(617,526){\makebox(0,0)[l]{p + $^7$Li}}
\put(567,526){\makebox(0,0)[l]{p + $^7$Li}}
%\put(869,526){\makebox(0,0)[l]{$^4$He + $^4$He}}
\put(819,526){\makebox(0,0)[l]{$^4$He + $^4$He}}
%\put(1058,526){\makebox(0,0)[l]{$^8$B}}
\put(1088,526){\makebox(0,0)[l]{$^8$B}}
\put(1310,526){\makebox(0,0)[l]{$^8$Be +$e^++\nu_e$}}
%\put(1058,450){\makebox(0,0)[l]{$^8$Be}}
\put(1088,450){\makebox(0,0)[l]{$^8$Be}}
\put(1310,450){\makebox(0,0)[l]{$^4$He + $^4$He}}
%\put(176.0,68.0){\rule[-0.200pt]{0.400pt}{368.095pt}}
\put(680,1520){\vector(1,0){126}}
\put(680,1290){\vector(1,0){95}}
\put(838,1443){\vector(0,-1){76}}
\put(428,1138){\vector(0,-1){77}}
\put(1058,1138){\vector(0,-1){77}}
%\put(428,985){\vector(1,0){95}}
\put(358,985){\vector(1,0){95}}
%\put(1090,985){\vector(1,0){63}}
\put(1040,985){\vector(1,0){63}}
\put(806,794){\vector(0,-1){115}}
\put(1184,794){\vector(0,-1){115}}
%\put(775,603){\vector(1,0){63}}
\put(735,603){\vector(1,0){63}}
\put(1216,603){\vector(1,0){63}}
%\put(775,526){\vector(1,0){63}}
\put(735,526){\vector(1,0){63}}
\put(1216,526){\vector(1,0){63}}
\put(1216,450){\vector(1,0){63}}
%\put(176,1596){\rule{1pt}{1pt}}
%\put(1436,1596){\rule{1pt}{1pt}}
\put(838,1237){\usebox{\plotpoint}}
\put(838.0,1138.0){\rule[-0.200pt]{0.400pt}{23.849pt}}
\put(428,1138){\usebox{\plotpoint}}
\put(428.0,1138.0){\rule[-0.200pt]{151.767pt}{0.400pt}}
\put(1058,870){\usebox{\plotpoint}}
\put(1058.0,794.0){\rule[-0.200pt]{0.400pt}{18.308pt}}
\put(806,794){\usebox{\plotpoint}}
\put(806.0,794.0){\rule[-0.200pt]{91.060pt}{0.400pt}}
\end{picture}
\end{table}

Of course $^7Be$-neutrinos can be suppressed more strongly, but 
their contribution to the total signal in the Homestake detector is small
($\sim 13\%$). The incompatibility of the Homestake and Kamiokande results 
was first recognized Bahcall and Bethe (1990): The contribution to 
the Homestake detector from B-neutrino flux only (taken from 
SuperKamiokande) is $2.71~SNU$, and thus it is larger than the total 
Homestake signal ($2.54~SNU$). 

It was demonstrated more accurately
(e.g. Bludman et al 1993, Castellani et al 1993, Hata et al 1994,
Berezinsky 1994, Bahcall 1994, Kwong and Rosen 1994, 
Degl'Innocenti et al 1995, Castellani et al 1997)
that comparison of Kamiokande and Homestake signals leave no place for 
$^7Be$-neutrinos. Recently, this conclusion was confirmed by Hata and 
Langacker (1997) for the new SuperKamiokande flux.

Neutrino oscillations, e.g. $\nu_e \to \nu_{\mu}$, easily solves the 
Homestake/Kamiokande conflict. Each case of $\nu_e \to \nu_{\mu}$ 
conversion results in full disappearance of a signal in the Homestake 
detector ($\nu_e+^{37}\!Cl \to ^{37}\!\!Ar+e$), while in the Kamiokande detector 
it still exists due to
$\nu_{\mu}+e \to \nu_{\mu}+e$ scattering. As a result the signal in 
Kamiokande should be stronger than in Homestake, as observed.

The deficit of beryllium neutrinos is seen not only from 
Homestake/SuperKamiokande conflict, but from comparison of the gallium 
signal with the signal from the SuperKamiokande or Homestake detectors
(see the references cited above and recent calculations by Hata and 
Langacker 1997). 
Comparison of 
signals from any two detectors results in null beryllium-neutrino flux 
if neutrino is standard. However, one can notice that inclusion of the 
gallium signal, needs the solar-luminosity sum rule, i.e. connection between 
$pp, Be$ and $B$ neutrino fluxes and solar luminosity, which is valid only 
for stationary Sun.

Thus, the deficit of the beryllium-neutrino flux, i.e. the strong suppression 
of this flux to the value much smaller than predicted in the SSMs, looks 
as the most serious problem of the astrophysical solution to the SNP. This 
problem
is naturally solved by elementary-particle solutions, where suppression 
of the fluxes is energy dependent. Beryllium neutrino flux will be directly 
measured by Borexino detector at Gran Sasso (Bellini 1996).

The second solar-neutrino problem is the deficit of the boron neutrinos.
For the last 4-5 years we thought that with extreme and correlated 
uncertainties in 
$pBe$-cross-section and in the central temperature $T_c$ this discrepancy 
can be eliminated. Now with the new data of SuperKamiokande and reduced
errors in $T_c$ due to helioseismic data (1.4\%), the deficit of boron neutrinos
is real again (Ricci et al 1997): the theoretical lower limit on boron 
neutrino flux is $2.2 - 3.5 \sigma$ higher than the SuperKamiokande flux.

In the light of all known physics included and confirmation by seismic data,
one cannot doubt any more, that the SSMs give {\em at least} the 
good approximation to the description of the sun. Then one can ask, whether 
there is a track from predictions of the SSMs to the observed solar 
neutrino flux, due to any (even unreasonable) change of the input parameters 
or due to ad hoc inclusion of the non-standard physics.

It is demonstrated that such track does not exist in case when these parameters 
are all nuclear cross-sections and central temperature $T_c$ (Berezinsky et 
al 1996).

The failure of inclusion of non-standard physics is reviewed in the book by 
Bahcall (1989). %\\*[135mm]
\begin{center}
\psfig{bbllx= 85pt, bblly=190pt, bburx=500pt, bbury=590pt,
file=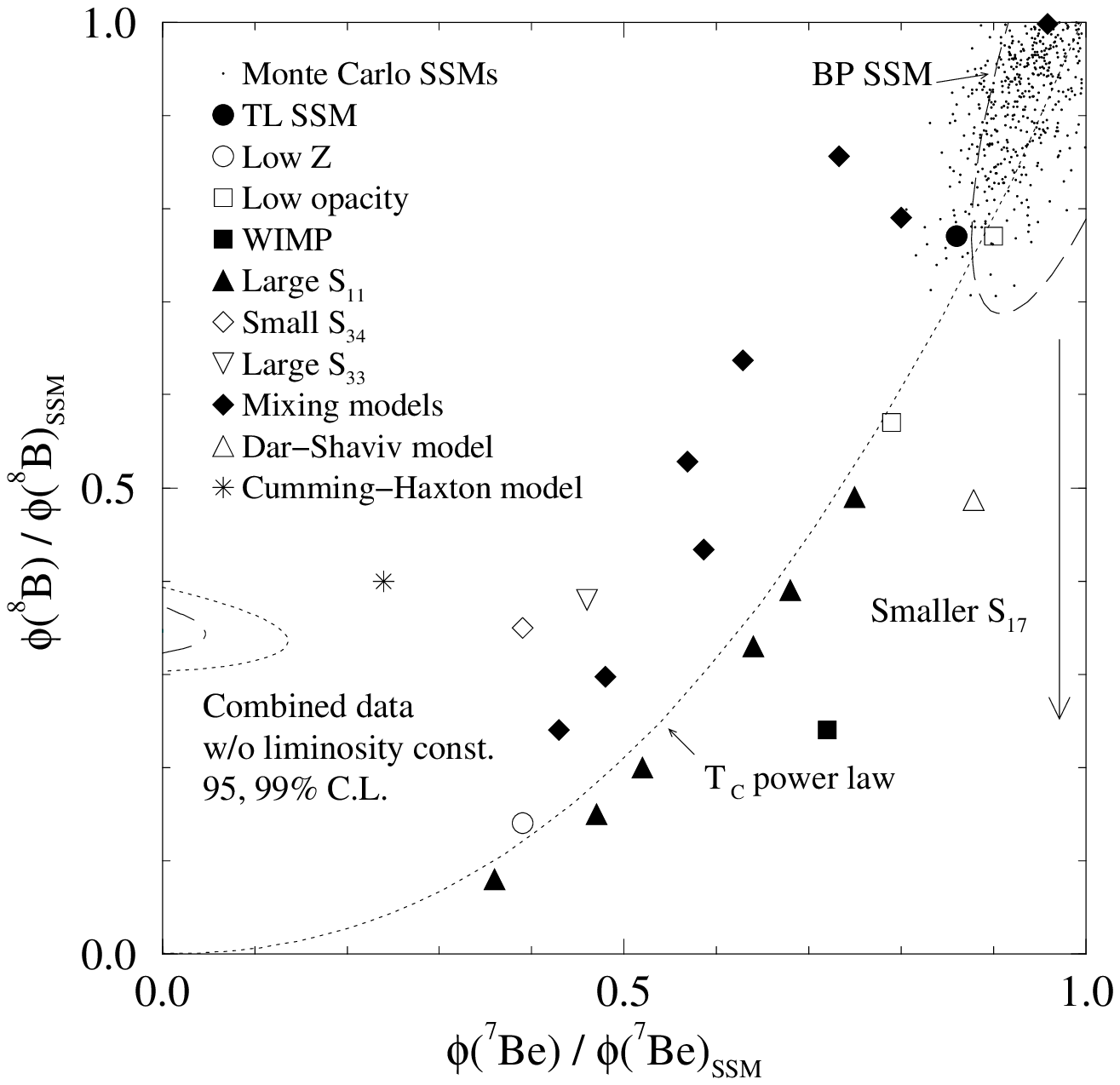, height= 12cm , clip=}
\end{center}
\noindent
Fig.1.  (from Hata and Langacker 1997). Is there a track from the region 
allowed by the SSMs (upper-right corner) to the region allowed by combined 
observational data (left corner)? The luminosity constraint is not imposed 
in calculations of the allowed region.

\newpage
 The modern status of this problem is clearly illustrated by 
Fig.1 from Hata and Langacker 1997  (note, that solar-luminosity sum rule
is not imposed here). In accordance with calculations by 
Berezinsky et al 1996, one can see that no track leads from the SSMs models
to the experimentally allowed region. The predictions of non-standard 
models are shown there as well. The Cumming-Haxton model (1996) reached  
the closest proximity to the allowed region. In this model the mixing in 
the inner core is {\em assumed}. Non-equilibrium $^3He$ coming to the center
,
produces $^4He$ through $^3He+^3\!\!He$, reducing thus production of boron 
and beryllium neutrinos. However, together with $^3He$ hydrogen gets 
in the inner core. It lowers molecular weight, increasing thus the sound 
speed. The calculations (Bahcall et al 1997) show that it exceeds 
the value allowed by seismic data.\\*[4mm]
\noindent
3. HELIOSEISMOLOGY\\*[4mm]
\noindent
As far as the SSMs are concerned we were very formal in the section 2, assuming 
they give only some approximation to the real picture inside the sun. In fact, 
these models are very well confirmed by the precise seismic measurements 
of sound speed and density profiles inside the sun.

Due to opposite forces of pressure and gravity, the sun can oscillate 
relative to its equilibrium configuration. The oscillations considered as 
small adiabatic perturbations, are described by usual set of equations 
for compressible selfgravitating gas. These equations give two solutions:
the high-frequency pressure modes (p-modes)and low-frequency gravitational 
modes (g-modes). Till now only p-modes are observed with typical periods 
from several minutes to 1 hour. The acoustic waves (p-modes) propagate 
inside the sun non-radially. Approaching the deep interior, the wave is 
deflected from radial direction due increasing sound speed and finally 
is reflected (effectively due to gradient of sound speed). On its way 
back it is reflected again from outer layers due to pressure gradient.  Thus 
acoustic waves are trapped in the solar cavity, forming the standing waves.  

The photosphere of the sun
is oscillating, forming a mosaic pattern, where various spots are 
moving with different velocities. These velocities can be measured due 
to e.g. Doppler shifts of optical lines. Thus the map of velocity field can 
be measured on the sun surface. The accuracy of velocity determination 
reaches $1~ cm/s$, though the typical one is $\sim 10~cm/s$. 

The oscillation modes are characterized by the spherical harmonic numbers 
$l$, $n$, and corresponding frequency $\omega_{l,n}$. The dependence on 
the azimuthal number $m$ is absent for spherically symmetric oscillations.
The data consist of the large set of frequencies for the
different numbers $l$ and $n$. The frequencies are typically in the range 
between $1000$ and $10000~ \mu Hz$ and the values of $l$ are between 
$l=0$ and $\sim 1000$ (see paper by M.Huber 1997 in these Proceedings for 
these and other details of helioseismology).     

The tremendous accuracy of frequency measurements ($\sim 10^{-6}$) is a basis 
for precise determination of solar parameters in helioseismology.

The acoustic wave with given $l$ is reflected at the distance $r_t$ inside 
the sun (Christensen-Dalsgaard 1996a):
\be
r_t=c(r_t)\sqrt{l(l+1)}/\omega_l,
\label{eq:r-t}
\ee
where $c$ is the adiabatic sound speed and $\omega_l$ is an angular
frequency. As follows from Eq.(\ref{eq:r-t}), the modes with large $l$ 
are reflected from large distance from the solar center, while modes with 
$l=0$ penetrate into center.

To {\em invert} the measured frequency pattern $\omega_{n,l}^{obs}$ into 
some physical values inside the sun (e.g. sound speed $c$ and density 
$\rho$ 
radial profiles) one needs a solar model as an auxiliary tool. Using this 
model  one calculates the difference of the frequencies 
$\delta \omega_{n,l}=\omega_{n,l}^{obs}-\omega_{n,l}^{mod}$. Using 
variational principal one can connect the variations of frequencies 
$\delta\omega_{n,l}$
with variations of the solar parameters $\delta c^2(r)$ and $\delta\rho(r)$ 
(see e.g. 
Gough and Thompson 1991, Christensen-Dalsgaard 1996a):
\be
\frac{\delta \omega_{n,l}}{\omega_{n,l}}= \int_0^Rdr K_{c^2}^{n,l}(r)
\frac{\delta c^2(r)}{c^2(r)}+ \int_0^R dr K_{\rho}^{n,l}(r)
\frac{\delta \rho(r)}{\rho(r)} + \frac{F_{surf}(\omega_{n,l})}{I_{n,l}},
\label{eq:inver}
\ee
where the last term in Eq.(\ref{eq:inver}) is added to account for  more 
complicated physics near the surface of the sun; $I_{n,l}$ are the calculated 
moments of inertia. The kernels of Eq.(\ref{eq:inver}) are calculated in the 
reference (auxiliary) model. The mathematical inversion of Eq.(\ref{eq:inver}) 
yields 
$c^2(r)$ and $\rho(r)$ in the solar interior. The success of this method 
depends on the choice of the reference model: the better is the model the 
smaller deflections $\delta c^2$ and $\delta \rho$ one obtains in the end.
Apart from this method there is another one, where less precise asymptotic 
formulae are used, but which does not need an auxiliary model (for a 
review see Turck-Chieze 1993).\\*[5mm]
\begin{center}
\epsfig{file=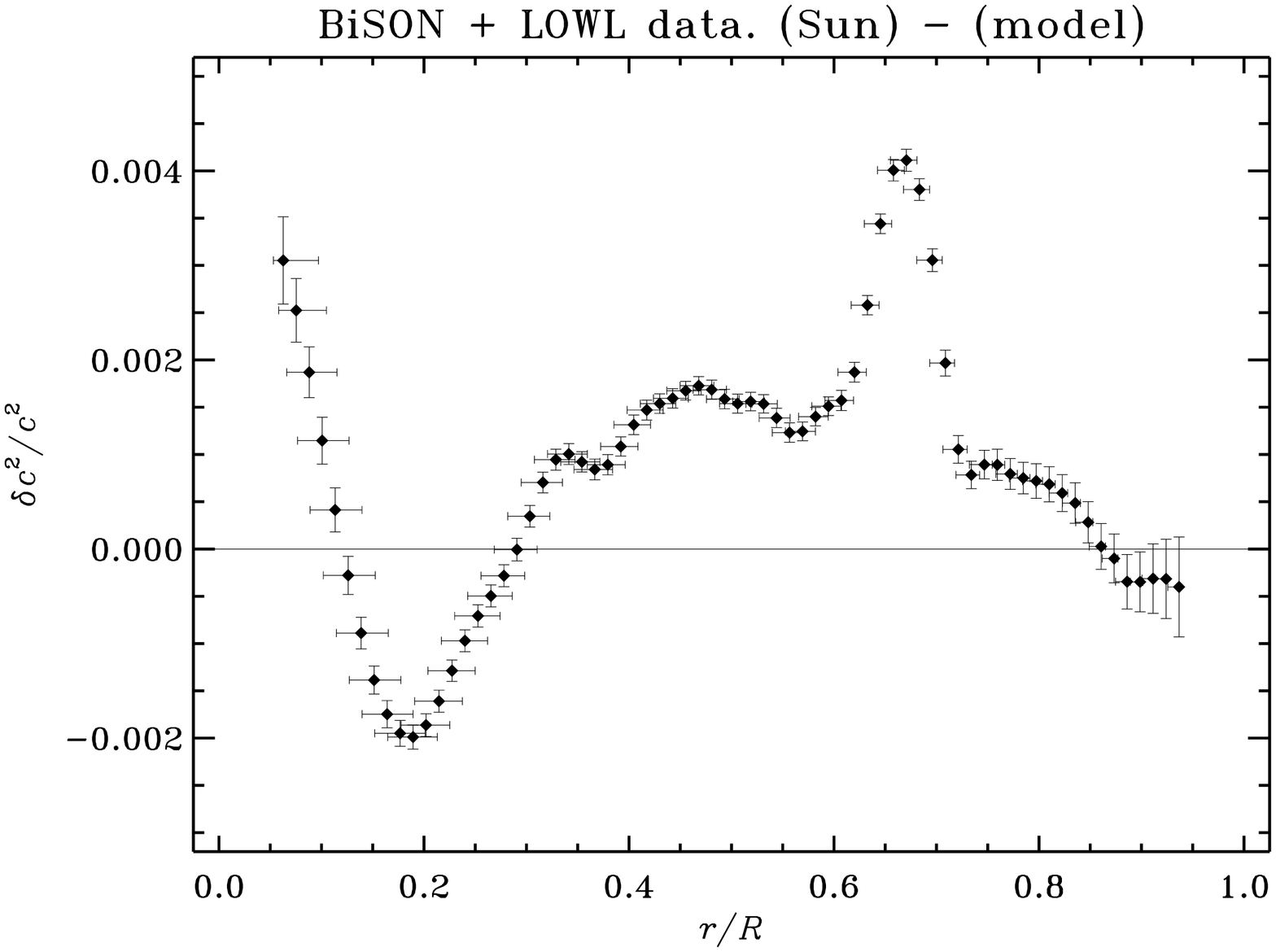,height=10cm,angle=90}
\end{center}
\noindent
Fig.2 (from Christensen-Dalsgaard 1996b). The seismic sound speed profile
compared with prediction of the reference model S  of Christensen-Dalsgaard et 
al (1996c). Shown 
are deflections of sound speed squared from the calculated value.\\*[5mm]
\begin{center}
\epsfig{file=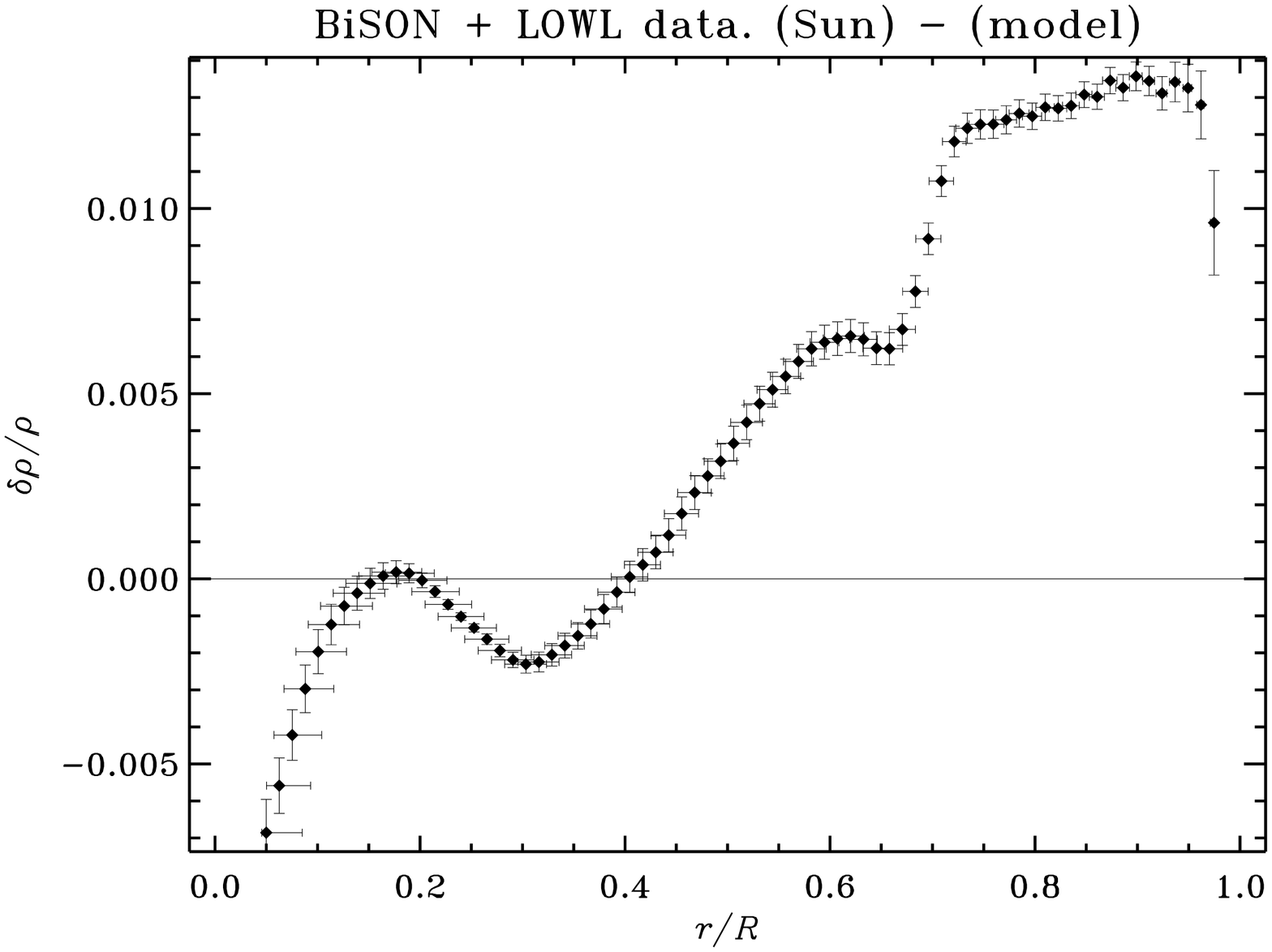,height=10cm,angle=90}
\end{center}
\noindent
Fig.3 (from Christensen-Dalsgaard 1996b). The seismic density profile
compared with prediction of the reference model S.

\newpage
In Fig.2 and 3 the sound speed and density profiles, as presented by 
Christensen-Dalsgaard (1996b), are shown. As a reference model is used 
the standard model S of Christensen-Dalsgaard et al (1996c) with the diffusion 
of elements and settling included. The seismic data from the BiSON network 
and LOWL detector (SOHO satellite) are used. The horizontal line gives 
zero deflections from the SSM. Apart from two irregularities at 
$r/R \approx 0.7$ and $r/R \to 0$, where $R$ is the solar radius, the 
agreement of $c^2$ (Fig.2) with the model calculated values is better than 
$0.2\%$.  For the density (Fig.3) the agreement at distances 
$r/R < 0.7$ is better than $0.7\%$. 

The most disturbing feature for the SNP in these results is increasing 
discrepancy 
between the SSM and observations in the inner core ($r/R \leq 0.1$): 
neutrino fluxes are produced mostly in the inner core.

A break-through in this problem was reached recently due to works 
by Dziembowski (1996) and Degl'Innocenti et al (1997a), and by Bahcall et al 
(1997). 

As was explained above, only low $l$ modes penetrate to the inner core.
Dziembowski (1996) and Degl'Innocenti et al (1997a) 
used in their analysis 33 modes with $ 0\leq l\leq 2$ and 
another 20 modes in the range of $l$ between 5 and 13. The new method 
of inversion was developed for the case of small number of frequencies, 
which was used 
with an additional assumption of smoothness of the functions.
The results for isothermal sound speed $u=p/\rho$, where $p$ is a pressure, 
is shown in Fig.4.%\\*[110mm]
\begin{center}
\psfig{bbllx= 65pt, bblly=360pt, bburx=540pt, bbury=670pt,
file=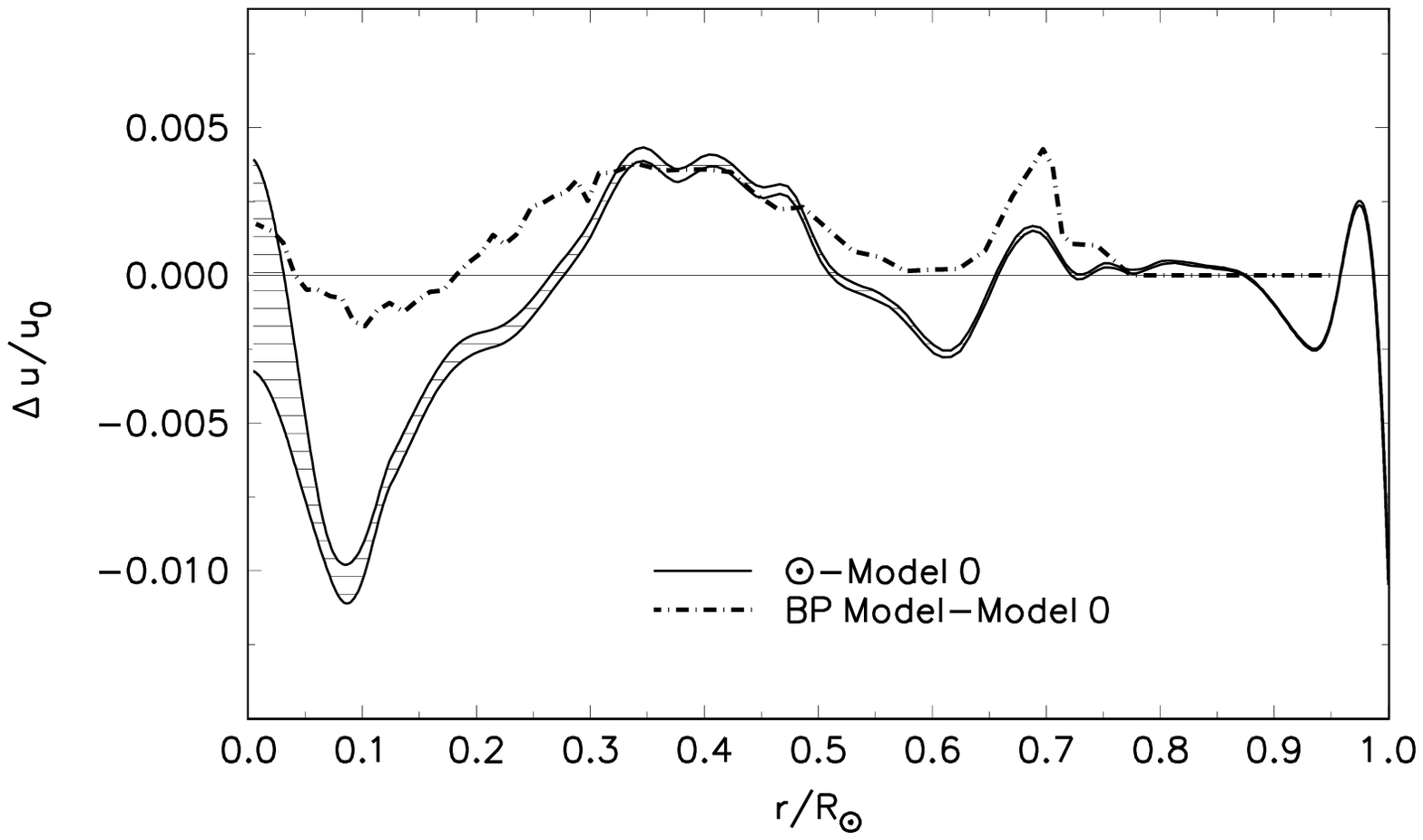, height= 11cm , clip=}
\end{center}
\noindent
Fig.4 (from Dziembowski 1996). Isothermal sound speed ($u=p/\rho$) profile
compared with the reference model of the author and with the SSM by 
Bahcall and Pinsonneault 1995. Shown is deflection of the sound speed 
$\Delta u$ from the calculated value.\\*[1mm]

One can observe from Fig.4 that increasing of the 
deflections towards the inner core is changed for 
better agreement at smaller distances. Apart from the reference model 0 of 
the authors, this conclusion is valid also for the Bahcall-Pinsonneault (1995)
model (the dash-dotted curve). However, the uncertainties of this result 
are rather large (Degl'Innocenti et al 1997a).

The excellent agreement of the Bahcall and Pinsonneault (1995) model with 
acoustic data in the inner core can be seen from analysis by Bahcall et al 
(1997). At $r/R \leq 0.2$ the agreement is better than $0.2\%$.

What is the status of the SSMs in the light of their confirmation by 
seismic data?

The accuracy of of the SSM predictions better than $1\%$ is actually more 
than is needed now for the calculations of solar-neutrino fluxes, because
of two reasons. First, the uncertainties in the cross-sections are much 
larger. Second, the solar-neutrino deficit is most probably a result of 
neutrino oscillations, and the parameters of these oscillations must be 
in the end measured directly.

On the other hand the SSMs are confirmed by seismic observations only by 
two physical quantities: $\rho(r)$ and $c_s(r)$, or $p(r)$. Could it be 
that the other parameters important for calculations of solar-neutrino
fluxes, e.g. temperature $T(r)$ and chemical composition $Y(r), Z(r)$ etc,
are given by the SSMs incorrectly? This question was raised by many people.

One may argue that this is unnatural possibility (see also Bahcall et al 1997). 
The adiabatic sound speed in fully ionized inner core,
$c=\sqrt{\gamma RT/\mu}$, where $\mu$ is a molecular weight, and 
therefore one has for the small variations
\be
\delta c/c=(1/2)(\delta T/T-\delta \mu/\mu).
\label{eq:c-s}
\ee
In the SSMs all three values in Eq.(\ref{eq:c-s}) are close to each other.
One can probably assume accidental fine-tuning (the compensation of large 
values in the 
rh-side of Eq.(\ref{eq:c-s})) at some particular distance $r$, but it is 
hard to imagine a model with such compensation at all $r$.

In principle, the temperature can be directly measured by the width of 
the $^7Be$-neutrino line (Bahcall 1993), but the energy resolution 
needed ($\sim 10^{-3}$) leaves this problem for the future technique.

One should not consider helioseismology just as auxiliary tool for solar 
neutrino physics. Its main task is to explore the finer structure of the 
sun than it is needed for the sake of the solar-neutrino physics. 
The seismic measurements give us clear indications to some phenomena not 
described by the SSMs. One can see in Fig.2 statistically very significant 
peculiarities at $r/R \sim 0.2$ and $r/T \sim 0.7$. The density profile 
(Fig.3) also exhibits the jump at $r/R \sim 0.7$, as well as small but 
regular deflection from the SSM density profile. It is interesting to note 
that both abovementioned irregularities occurs at distances where the SSMs
predict the large composition gradients (Christensen-Dalsgaard 1996b). It 
can result in the mixing, which naturally produces (through the change of 
molecular weight) irregularities in the sound speed profile seen in Fig.2.
However, the large mixing, which could affect the solar-neutrino fluxes,
are ruled out (Bahcall et al 1997).

The SSMs are based on the following basic assumptions (see Bahcall 1989, 
Castellani et al 1997):

(i) The solar energy is generated by nuclear reactions (the calculations 
show that $98\%$ of the energy is supplied by the $pp$-chain and $2\%$ by
the CNO-cycle).

(ii) The sun evolves in hydrostatic equilibrium, with gravitational 
attraction balanced by the pressure. The equation of state, which is 
employed to calculate the pressure, includes many fine effects, and probably
can be considered as reliable.  
Instabilities are assumed not to be 
present. The effects due to sun rotation are neglected.

(iii) The energy transport is radiative in the solar core and convective 
outside it. Convection is treated according to rather simplified 
mixing-length theory (Christensen-Dalsgaard 1996a). The collective plasma 
effects (Tsytovich et al 1995) are neglected for radiative opacity.

(iii) When the sun first entered  the Main Sequence, it was highly 
convective  and therefore uniform in composition. It implies that the 
present chemical composition of its surface is the same as the initial 
chemical composition of the sun.

(iv) The diffusion of heavy elements and gravitational settling are 
included in the calculations: the heavy nuclei ($He$ and heavier elements)
diffuse to the center, due to excessive gravitational force, faster than 
hydrogen.

The model of the sun is evolved until it reaches the present luminosity.
The calculated solar radius and age are in agreement with the observations.
The most uncertain input parameters in these calculations are heavy element 
abundancies and calculated opacities.  
 
There are several minor effects which should be included now in the 
solar models, due to small discrepancies of the SSMs with seismic data.
Several kinds of instabilities can result in the small mixings.
Most widely discussed is one, driven by low $n$ and $l$ g-mode oscillations,
as was suggested first by Dilke and Gough (1972). Additional mixing can be 
caused by rotationally induced instabilities. Some mixing can occur beneath 
the convection zone. For the discussion of several effects of "non-standard" 
physics see Turck-Chieze et al (1993).

Can neutrino fluxes be calculated directly from seismic data? Apparently 
not, because for such calculations one must know the profile of chemical 
composition, which is given by some model of solar evolution. Ricci et al 
(1997) suggested an approach in which solar-neutrino fluxes are directly 
constrained by helioseismic data. This approach is not model-independent,
but a wider class of models in comparison with the SSMs are considered.
These models are based on the same equilibrium and evolution equations as 
the SSMs, but they allow the arbitrary choice of some input parameters.
The correct models are chosen in the end due to seismic constraints.
These models are called Helioseismically-Constrained Solar Models (HCSM).
The most trouble-some parameters of the SSMs are radiative opacity
$\kappa$ and the fraction of heavy elements $Z/X$, with relatively large 
errors. In the HCSM the arbitrary values of these parameters
are used. As seismic "observables" three independent 
physical quantities, determined most accurately  by seismic observations, 
are used, namely, the photospheric helium abundance, $Y_{ph}$, the depth 
of convective envelope, $R_b$, and the density, $\rho_b$, at the bottom of 
convective envelope. The central temperature calculated in this model is
\be
T_{HCSM}=1.58\times 10^7~K
\ee
If all errors involved in the calculations are summed quadratically,
the uncertainties in temperature determination is 
$(\Delta T/T)_{HCSM}=\pm 0.5\%$, to be compared with the similarly estimated 
uncertainties in the SSMs $(\Delta T/T)_{SSM}=\pm 1.7\%$. However, in  
the results cited below, we use more cautious (though less reasonable) 
approach, when all errors involved are summed up linearly. In this case we 
have $(\Delta T/T)_{HCSM}=\pm1.4\%$ and $(\Delta T/T)_{SSM}=\pm 2.7\%$.

Neutrino fluxes as calculated in several SSMs and the HCSM are given in 
Table 3., where BP95 refers to Bahcall and Pinsonneault (1995), FR96 and 
FR97 - to Degl'Innocenti et al (1997b), and JCD - to model S of 
Christensen-Dalsgaard et al (1996c). Not surprisingly, the HCSM predicts 
the neutrino fluxes in agreement with the SSMs, but with reduced 
errors.\\ 
\begin{table}[h]
\caption[abc]{
Predictions for neutrino fluxes and signals in the Cl and Ga detectors 
from the SSM and the HCSM. Uncertainties corresponding
 to $(\Delta T/T )_{HCSM}=\pm 1.4\%$ 
are shown (first error) together with those from
 nuclear cross sections (second error).
}
\center{\begin{tabular} {ll | c c c c c| c c c c c }
\hline
 & &\multicolumn{4}{c}{SSM} && \multicolumn{4}{c}{$~~~~~~~~$HCSM}$~~~~~~~~~$\\
 & & BP95 & FR97 & FR96  & JCD  &&  \\
\hline
$\Phi_{Be}$ &$[10^9$/cm$^{2}$/s$]$ & 5.15 & 4.49 & 4.58 & 4.94 &&  
4.81$\pm 0.53 \pm 0.59$ & \\
%$[10^9$/cm$^{2}$/s$^{1}]$ &&&&      &&&&\\
 $\Phi_B$ &$[10^6$/cm$^{2}$/s$]$ & 6.62 &  5.16& 5.28 & 5.87 && 5.96$\pm 1.49 
\pm 1.93$ &  \\
%$[10^6$/cm$^{2}$/s$^{1}]$ &&&&&&&&&\\
Cl &$[$SNU$]$& 9.3 & 7.3 & 7.5 & 8.2 && 8.4$\pm 1.9 \pm 2.2$   &\\
%$[$SNU$]$ & &     &     &     &&        &      &     &  &\\
Ga &$[$SNU$]$&137 &128& 129& 132 &&  133$\pm 11 \pm 8$ &\\
\hline
%$[$SNU$]$ & &   &    &     &&     &     &     & &
\end{tabular}}
\label{tabsig}
\end{table}
%\vskip 3mm
Note, that even when we sum errors linearly (the first errors 
in column HCSM), they are still less than the errors from nuclear 
cross-sections (the second error in the same column).
The contradiction of these predictions with the observed fluxes (Table 1)
demonstrates the conflict  between helioseismologically-constrained solar 
models and the  standard neutrino.\\*[3mm]

\noindent
4. NUCLEAR REACTIONS\\*[3mm]
\noindent
Uncertainties in cross-sections of nuclear reactions cannot solve the 
SNP (Berezinsky et al 1996), but these uncertainties cause now the largest 
errors in the prediction of solar-neutrino fluxes. The status of nuclear 
reactions in the sun was recently reviewed by Castellani et al (1997). 
The new, very recent progress is connected with measurements of 
$^3He+^3\!\!He$ cross-section in the LUNA experiment at Gran Sasso and 
with detailed calculations of the plasma effects in nuclear reactions 
(Gruzinov and Bahcall 1997 and  Brown and Sawyer 1997). 
These effects include in particular electron screening and related 
problem of electron capture by $^7Be$. These calculations confirm
the previous results within accuracy needed for 
calculations of the solar-neutrino fluxes.

As was already mentioned in Introduction, the cross-section of 
$^3He+^3\!He \to ^4\!\!He+2p$ reaction could be of crucial importance 
for the SNP. If this cross-section were large at the energy of the Gamow 
peak ($\sim 20~keV$), e.g. due to presence of the narrow resonance at this 
energy, the production of $^4He$ in the sun would go mainly through the 
chain I (see Table 2), and thus production of beryllium and boron neutrinos
would be suppressed. The cross-section of this reaction as a function of 
energy is shown in Fig.5 (Arpesella et al 1996). The recent LUNA data 
(filled circles) correspond 
to  the solar Gamow peak and coincide well with extrapolation from 
higher energies in case of screening.%\\*[120mm]
\begin{center}
\psfig{bbllx= 50pt, bblly=230pt, bburx=545pt, bbury=600pt,
file=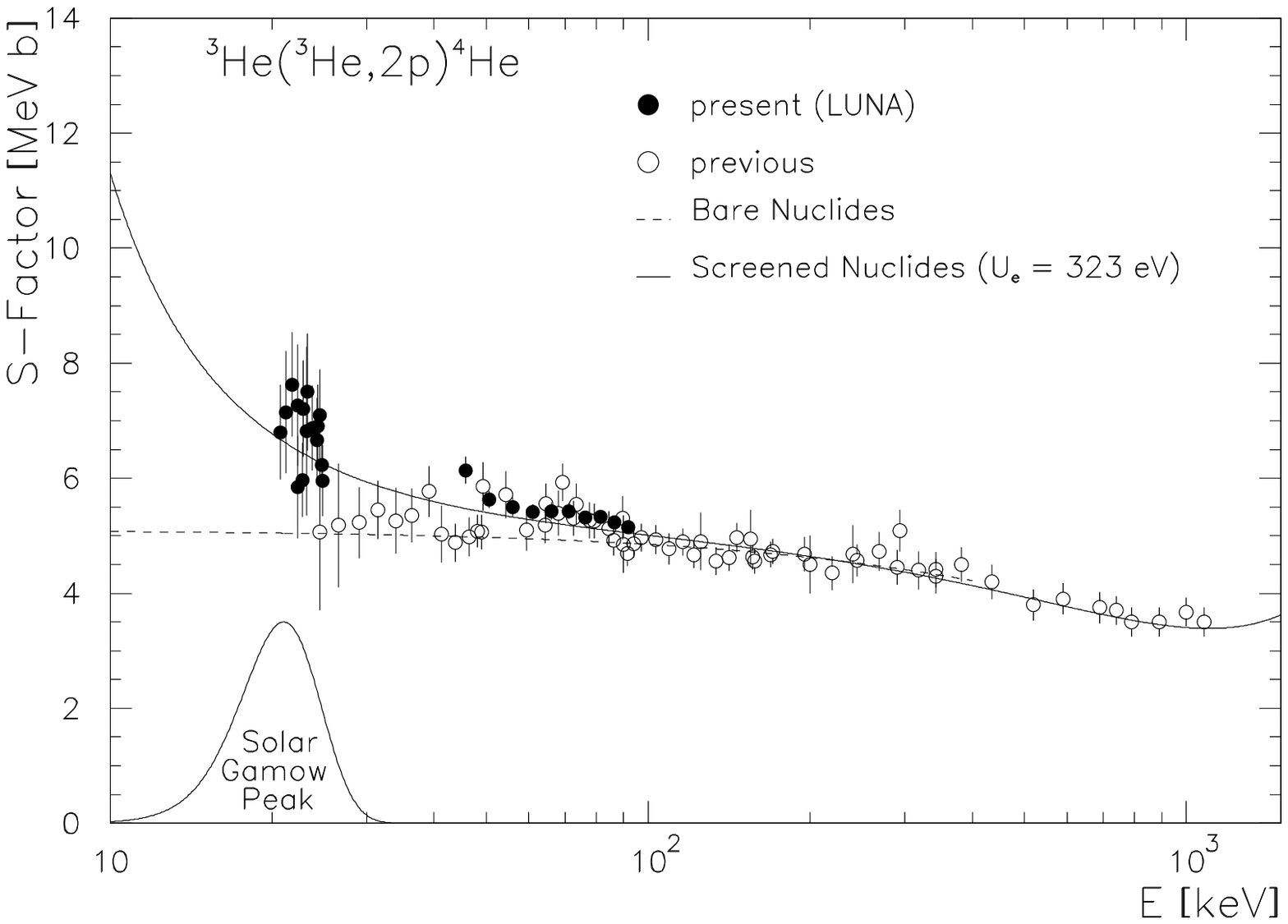, height= 12cm , clip=}
\end{center}
\noindent
Fig.5 (from Arpesella et al 1996). The compillation of 
astrophysical factors for $^3He+^3\!\!He$ cross-section. The LUNA data are 
shown by filled circles at the energy of the Gamow peak.\\*[1mm]

The most important reaction for boron neutrino production
$p+^7\!\!Be \to ^8\!\!B+\gamma$ has the largest uncertainties in the 
cross-section (see Castellani et al 1997). Most authors use for model 
calculations the astrophysical factor at the energy of the Gamov peak 
$S_{17}(0)= 22.4 \pm 2.1~eV b$, while much lower value, $16.7\pm3.2~eV b$
is obtained from Coulomb dissociation of $^8B$-nuclei.

The measurement of this cross-section at energy down to $E_{cm}=30~keV$
is planned for the second phase (1998 - 2002) of the LUNA experiment.
It is one of the reactions, which will be measured at the second phase.
The others are $p+D \to ^3\!\!He + \gamma$,
$D+^3\!\!He \to ^4\!\!He+p$ 
$^3He+^4\!\!He \to ^7\!\!Be+\gamma$, and 
$p+^{14}\!\!N\to ^{15}\!\!O+\gamma$ etc.\\*[4mm]
\noindent
5. ELEMENTARY-PARTICLE SOLUTIONS\\*[4mm]
\noindent
There are three elementary-particle solutions which can explain the 
results of solar-neutrino experiments: MSW effect (Mikheev and Smirnov
1986a, Wolfenstein 1978), vacuum oscillations (Pontecorvo 1957) and 
Resonant Spin-Flavor Precession, RSFP (Akhmedov 1988, Lim and 
Marciano 1988). In all these solutions electron neutrinos, born in
the sun, are converted into other neutrino states (muon neutrinos, 
sterile neutrinos or muon antineutrinos) and thus do not interact in the 
neutrino detectors, or interact weakly through neutral current effects  
(e.g. $\nu_{\mu}+e \to \nu_{\mu}+e$). The survival 
probability for $\nu_e$ depends on neutrino energy and actually this 
property allows to explain the different neutrino deficit in the 
solar-neutrino experiments. The maximum suppression must approximately 
correspond to the energy of beryllium neutrinos. 

In case of the MSW and RSFP, the conversion of $\nu_e$ into another neutrino 
flavor occurs {\em inside} the sun due to effects of matter oscillation.
Effect of  vacuum oscillation takes place on the way from the sun 
to earth.
 
All three solutions are characterized by the difference of neutrino masses 
squared , $\Delta m^2$ and by mixing angle, $\theta$ between two neutrino mass
states $\nu_1$ and $\nu_2$, which form neutrino flavor state, e.g. 
$\nu_e$ or $\nu_{\mu}$. The 
energy shape of suppression curve for neutrino fluxes is determined by 
the pair of values ($\Delta m^2, sin^2 2\theta$) and therefore the 
different points in  $\Delta m^2 -  sin^2 2\theta$ plane give the 
different predictions for the signal in the neutrino detectors. It is not 
trivial that there are regions in this plane, which explain the 
results of all four solar-neutrino experiments. It is easy to understand 
that for the different regions the distortion of the produced neutrino 
{\em spectrum} is different.

Let us start with vacuum oscillations.

In case of two neutrino flavors, $\nu_e$ and $\nu_{\mu}$ are superposition 
of two mass states $\nu_1$ and $\nu_2$ with masses $m_1$ and $m_2$, 
respectively:
\be
|\nu_e>=\cos\theta|\nu_1>+\sin\theta|\nu_2> \\
\label{eq:mix}
\ee
\be
|\nu_{\mu}>=\cos\theta|\nu_2>-\sin\theta|\nu_1>
\ee
Here and everywhere else $\theta$ is the vacuum mixing angle.
As a result of $\beta$ decay in the sun $\nu_e$-neutrino with a fixed energy 
$E$ is produced, and therefore the states $\nu_1$ and $\nu_2$ have the 
different momenta $p_1 \approx E - m_1^2/2E$ and $p_2\approx E- m_2^2/2E$.
At the distance $r$ from the source the states $\nu_1$ and $\nu_2$ obtain
different phases $\exp(ip_1r)$ and $\exp(ip_2r)$. Thus, $\nu_e$ from 
Eq.(\ref{eq:mix}) becomes
\be
|\nu(r)>=\exp(ip_1r)\cos\theta|\nu_1>+\exp(ip_2r)\sin\theta|\nu_2>,
\ee
which now has admixture of $\nu_{\mu}$. The probability, that performing 
an experiment one finds this neutrino as the muon neutrino is easy to 
calculate as 
\be
P_{\nu_e \to \nu_{\mu}}(r)=|\nu(r)><\nu_{\mu}|^2=\sin^2(2\theta) 
\sin^2\left(\frac{\Delta m^2}{4E}r\right),
\label{eq:prob}
\ee
where $\Delta m^2=m_1^2-m_2^2$ and $l_v=4\pi E/\Delta m^2$ is the vacuum 
oscillation length.

For the recent detailed discussion of vacuum oscillation solution for the 
SNP see Krastev and Petcov (1996) and Hata and Langacker (1997).

The regions in $\Delta m^2 - \sin^22\theta$ plane which explain all 
observational data and include the constraint due the Kamiokande energy 
spectrum, are shown in Fig.6 (Hata and Langacker 1997). One can see that 
only large-angle solutions are allowed, with very small
$\Delta m^2 \approx (5 - 8)\cdot 10^{-10}~eV^2$.%\\*[140mm]
\begin{center}
\epsfig{file=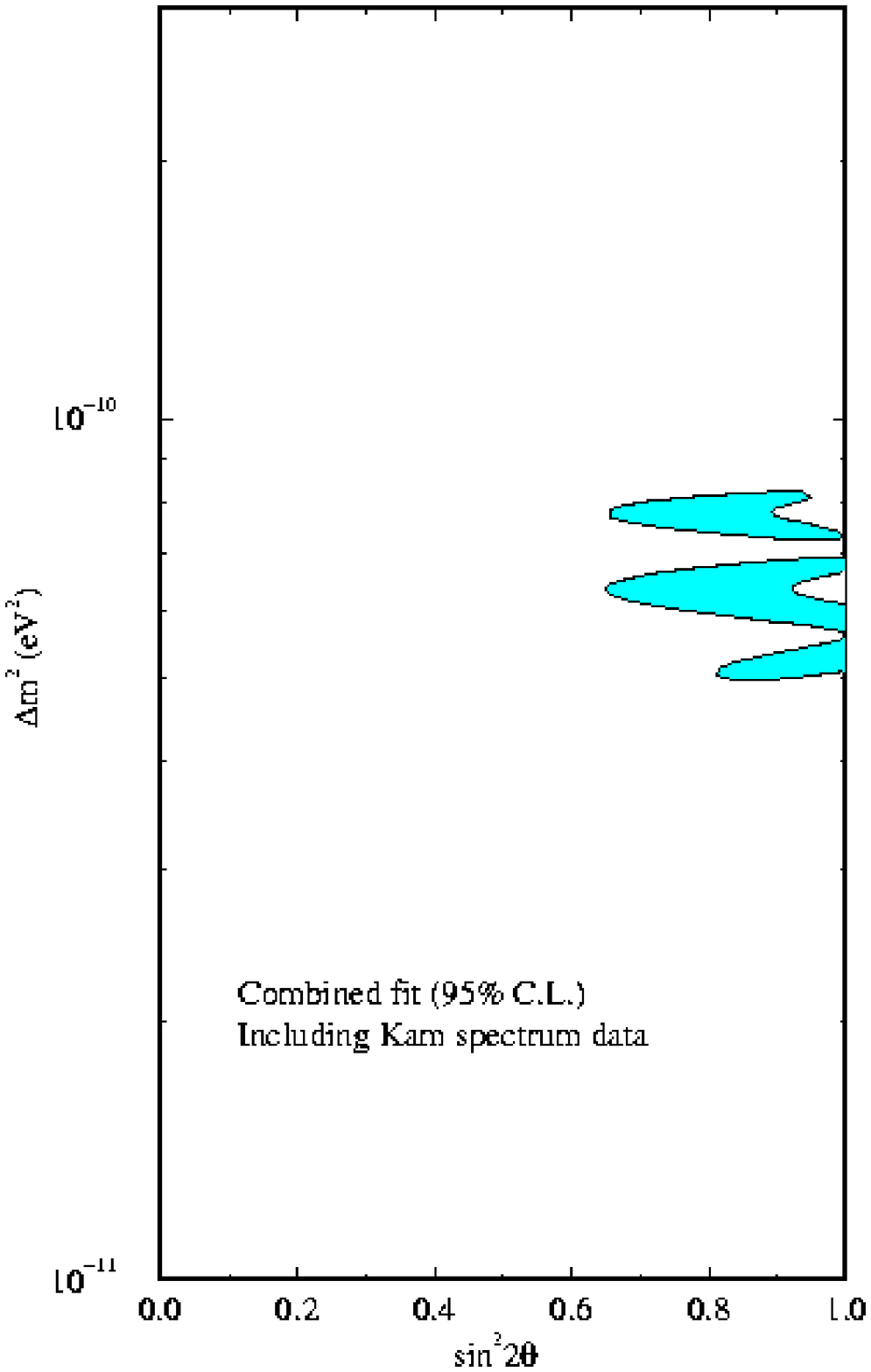,height=14cm}
\end{center}
\noindent
Fig.6 (from Hata and Langacker 1997). Vacuum oscillations: The regions 
allowed for explanation of all solar-neutrino data including the 
Kamiokande energy spectrum.\\*[1mm]  

The MSW effect describes the matter enhanced oscillations. It has a 
resonant character and for energies of neutrinos at interest typically 
occurs in the narrow layer, 
$\Delta R \sim 0.01 R_{\odot}$, at the distance $R\sim 0.1 R_{\odot}$
from the center of the sun. 

One can give the qualitative explanation of the adiabatic MSW effect. 

Let us consider the case when electron neutrino in vacuum is almost 
the light state $\nu_1$, while the muon neutrino is almost the heavy 
state $\nu_2$. Let both neutrinos have the same momentum $p>>m$. Then the 
energy of each neutrino is $E_1=p+m_1^2/2p+V_1$ and $E_2=p+m_2^2/2p+V_2$,
where $V_1$ and $V_2$ are matter-induced potential energies of neutrinos, 
which are determined by scattering of neutrinos in matter. The difference 
of neutrino energies at equal momentum is $\Delta E=\Delta m^2/2p +W(r)$,
where $W(r)=V_2-V_1$ is the difference of the potentials, determined by the 
different scattering of $\nu_e$ and $\nu_{\mu}$ on electrons: in the 
$\nu_e$ case it is provided by charged and neutral currents, while for 
$\nu_{\mu}$ -  
only by the neutral current. The explicit calculations give 
$W(r)=-\sqrt{2}G_Fn_e(r)$, where $G_F$ is the Fermi constant and $n_e(r)$ 
is the density of electrons inside the sun. Therefore, we obtain
$$
\Delta E = \Delta m^2/2p - \sqrt{2}G_Fn_e(r).
$$ 
In more adequate treatment, the difference of the diagonal terms of the 
Hamiltonian plays the role of $\Delta E$, and one has
\be
\Delta H = \cos 2\theta \Delta m^2/2p - \sqrt{2}G_F n_e(r)
\label{eq:mass}
\ee
Locally, $\Delta E$ (or $\Delta H$) is interpreted as due to difference in 
neutrino masses. Then at $r \to \infty$ when $n_e=0$, $\Delta E>0$,
i.e. muon neutrino is heavier than electron neutrino. At $r \to 0$,
when second term in rhs of Eq.(\ref{eq:mass}) dominates, electron neutrino 
is heavier. If density $n_e(r)$ changes slowly, adiabatic approximation 
holds, and electron neutrino propagating from the solar core outside,
remain on the heavy state trajectory. It means that it leaves the sun as 
muon neutrino. This conversion mainly occurs, when $\Delta H \to 0$.
According to Eq.(\ref{eq:mass}) it determines the critical density $n_c$, 
when conversion has the resonant character:
\be
2\sqrt{2}G_F\,p\,n_c= \Delta m^2 \cos 2\theta_v
\label{eq:res}
\ee

 The present status of the MSW solution is illustrated by Fig.7 (Bahcall and 
Krastev 1997). The SSM used in the calculations is the one by Bahcall and 
Pinsonneault (1995). At $95\%$CL there are two allowed regions, given by 
the {\em small angle} solution around the point of local $\chi^2$ minimum
$\Delta m^2=5\cdot 10^{-6}~eV^2$ and $\sin^22\theta =9\cdot 10^{-3}$, and 
the {\em large-angle} solution with the $\chi^2$-min point 
$\Delta m^2=1.2\cdot 10^{-5}~eV^2$ and $\sin^22\theta=0.6$. At $99\%$ CL 
the third large angle solution appears.%\\*[120mm]
\begin{center}
\epsfig{file=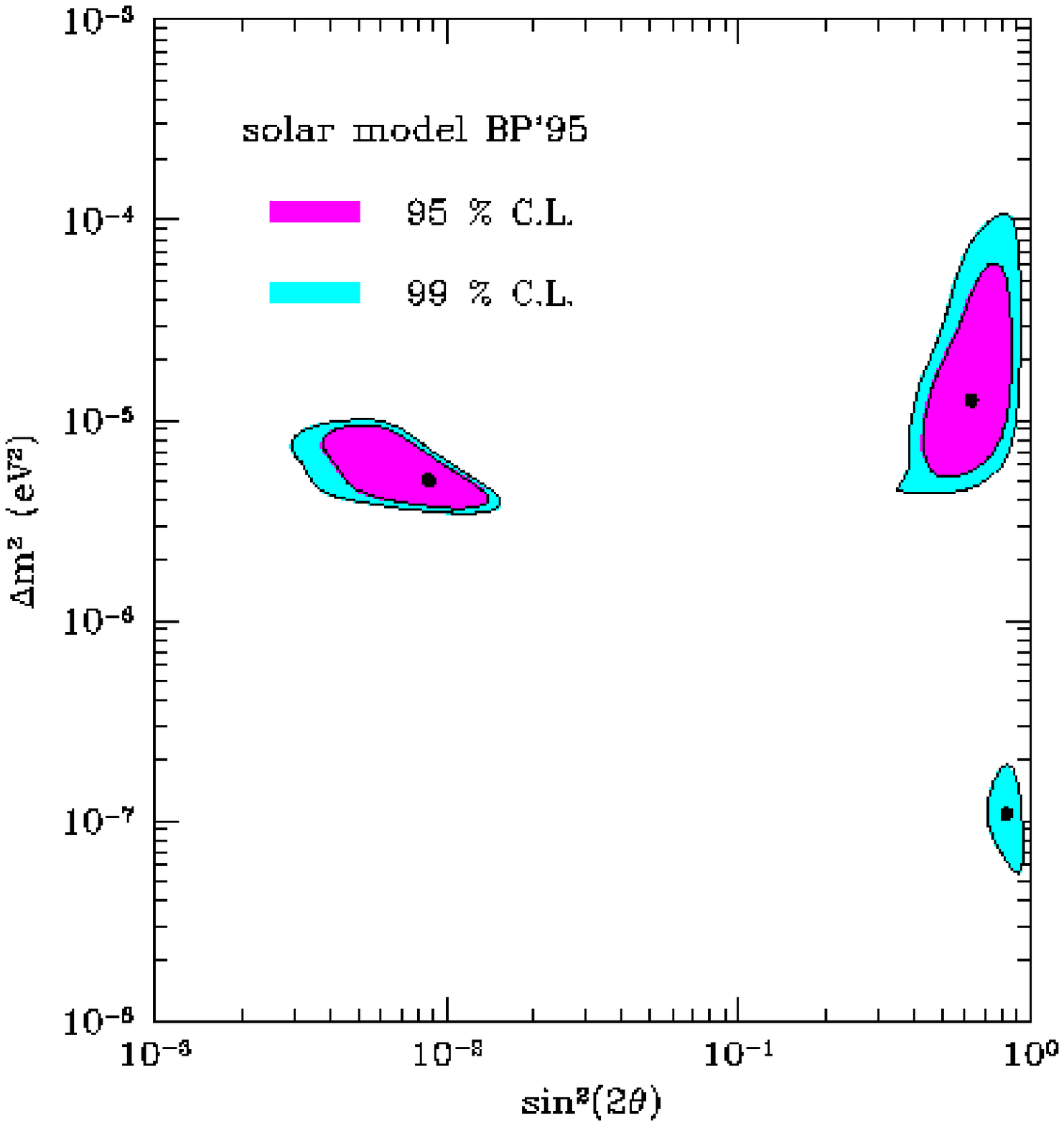,height=12cm}
\end{center}
\noindent
Fig.7 (from Bahcall and Krastev 1997). MSW conversion: The regions allowed 
for explanation of all solar-neutrino data.\\*[1mm]

The suppression curves, i.e. the survival probability $P(\nu_e\to \nu_e)$
as a function of neutrino energy, are given in Fig.8 (Berezinsky 1995)
for the small angle MSW solution with the different $\sin^22\theta$.
One can see that for the central point of the solution 
with $\sin^22\theta=9\cdot 10^{-3}$
the spectrum is very strongly distorted in the energy region 
$6 - 14~MeV$, where the measurements of SuperKamiokande and SNO will be 
soon available. However, as was noted by Krastev and Smirnov (1994) and 
by Berezinsky et al (1994) (see Hata and Langacker 1997, Fig.18, for recent 
calculations) the small reduction of calculated boron-neutrino flux,
for example due to diminishing $S_{17}$, shifts the small-mixing solution 
towards the smaller mixing angles. In particular the value 
$sin^22\theta \approx 1\cdot 10^{-3}$ is possible due to present 
uncertainties in $S_{17}$. In this case one can see from Fig.8 that 
for the energies $E\geq 6~MeV$, accessible for SuperKamiokande and SNO,
the distortion of energy spectrum is small.%\\*[110mm]
%\psdraft
\begin{center}
\epsfig{file=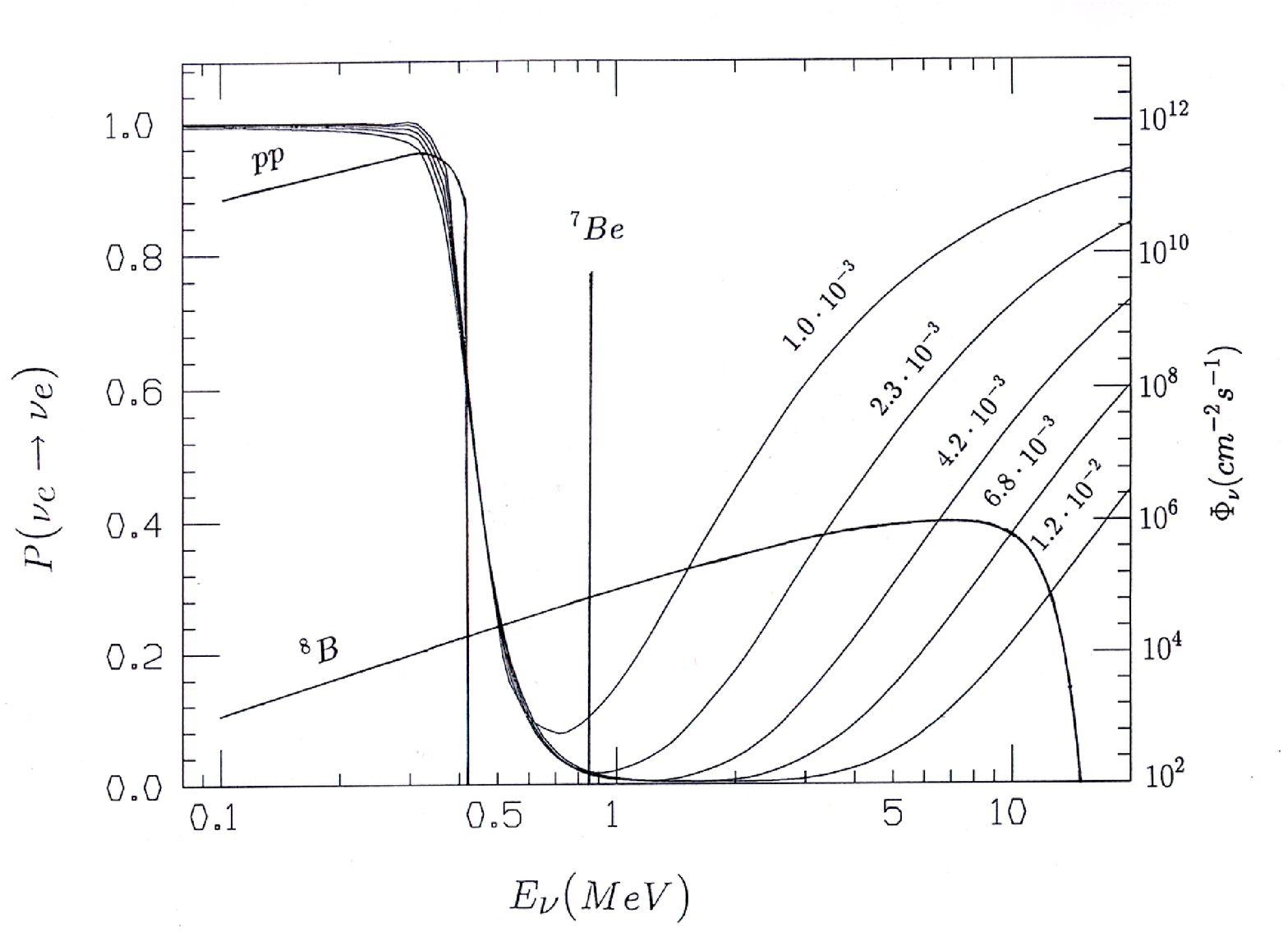,height=11cm}
\end{center}
\noindent
Fig. 8 (from Berezinsky 1995). The suppression factor $P(\nu_e\to \nu_e)$ 
for the MSW small-angle 
solution for different values of $sin^22\theta$ (indicated on the curves).
On the right vertical axis the neutrino fluxes are shown.\\*[1mm]

The Resonant Spin-Flavor Precession (RSFP) describes two physical effects
working simultaneously: the spin-flavor precession, when neutrino spin 
precesses around magnetic field, changing simultaneously neutrino flavor, and 
the resonant, density-dependent effect, which produces difference in potential 
energy of 
neutrinos with different flavors (a la the MSW effect). This complicated 
transition occurs in the external magnetic field due to presence of 
non-diagonal neutrino magnetic moments. The RSFP was recognized 
simultaneously by Akhmedov (1988) and Lim and Marciano (1988). For excellent 
recent review see Akhmedov (1997).

This theory had a predecessor. The neutrino spin precession was studied by 
Voloshin, Vysotsky and Okun (1986). The precession of neutrino magnetic 
moment around magnetic field converts left electron neutrino $\nu_{eL}$
into sterile right component $\nu_{eR}$, suppressing thus 
$\nu_e$-flux. However, the suppression effect in this case is 
energy-independent and thus contradicts to the recent solar-neutrino data.
Voloshin, Vysotsky, Okun (1986) and Barbieri and Fiorentini (1988) 
included the matter effects in the spin precession, and Schechter and Valle
(1981) discovered spin-flavor precession. The latest results of 
solar-neutrino experiments can be explained only by the RSFP, because 
only this type of precession  give the energy-dependent 
suppression factor. 

 In case of Majorana neutrino the RSFP induces 
the transition $\nu_{eL}$ to $\bar{\nu}_{\mu R}$, i.e.
to another active neutrino component (it can scatter off the electron). 
The suppression factor was calculated e.g. by Akhmedov et al (1993) and by 
Lim and Nunokawa (1995). It has 
different shape in comparison with the one for the MSW or vacuum-oscillation 
solution. At large neutrino energies it tends to 1/2, compared with $1$  
for vacuum oscillations and the MSW.
It means that the total suppression of SuperKamiokande 
signal is always larger than in the case of the MSW solution. It looks like 
a difficulty for this solution. However, Lim and Nunokawa (1995) found 
that the RSFP can explain all solar-neutrino data for the following range 
of parameters: neutrino magnetic moment $\mu=1\cdot 10^{-11}~\mu_B$,
magnetic field in the range $25 - 130~kG$ and $\Delta m^2$ in the range
$7\cdot 10^{-9} - 2\cdot 10^{-7}~eV^2$. The neutrino magnetic moment 
$\mu = 1\cdot 10^{-11} \mu_B$ is in conflict with the observations of 
the helium flashes of the red giants ($\mu < 3\cdot 10^{-12}~\mu_B$,
Raffelt (1990).\\*[3mm]
\noindent
6. SIGNATURES OF ELEMENTARY-PARTICLE SOLUTIONS\\*[3mm]
\noindent
The common signature of all three elementary-particle solutions is 
distortion of neutrino energy spectrum. It seems very unlikely that 
spectrum distortion will not be found in SuperKamiokande 
and SNO during the next 3-5 years. The energy spectrum will be 
precisely measured in the future experiments, ICARUS (ICARUS Collaboration
1994) and HELLAZ (Ypsilantis 1992 and references therein). Another 
common feature of all three solutions is that maximum of suppression 
curve corresponds to the energy of $^7Be$ neutrinos. Borexino at Gran Sasso
will be able to observe this effect (see Bellini 1996 and references 
therein). It is worth to note that Borexino will detect not only 
$\nu_e$-flux, but also another final state of oscillation, if it is active
(e.g. $\nu_{\mu}, \bar{\nu}_{\mu}$ etc). The recent suggestion by 
Raghavan (1997) of measuring fluxes of sub-MeV $\nu_e$-neutrinos due 
to inverse beta decays on $^{176}Yb$, $^{160}Gd$ and $^{82}Se$ looks like 
an excellent supplementary experiment for the Borexino.

It is much more difficult to distinguish experimentally between the three 
abovementioned solutions. The suppression curves for the MSW and vacuum 
oscillation solutions are similar in all cases and in some cases they are 
almost identical. There are some additional effects which can distort
the shape of suppression factor (e.g. the density fluctuations in case of 
the MSW effect - see Nunokawa et al 1996). What is more important,  
the small difference in the spectral shape cannot be considered as a final 
proof of discovery, for example, the MSW effect. It can be taken as an 
indication, but not more.

Fortunately, there are clear signatures for each solution.

For MSW-solution this signature is the day-night effect, caused by 
neutrino matter-oscillations in the earth (Mikheev and Smirnov 1986b, 
1986c   
and references therein). The neutrino spectrum from the sun will be 
additionally distorted in the {\em night} by the MSW effect in the earth.
The effect is two-fold: $\nu_e$-neutrinos are converted with some 
probability into $\nu_{\mu}$
(we limit ourselves by these two flavors only), and $\nu_{\mu}$, produced     
in the sun due to the MSW effect, will be partly regenerated in the earth 
back to $\nu_e$. These effects depend on the part of the earth neutrinos 
cross. The effect is stronger when neutrinos cross the earth core,
and weaker when they pass through the mantle only. The detailed calculations 
for the day-night effect relevant for superkamiokande were recently 
performed by Maris and Petcov (1997) (see also the references for the 
early calculations there). The value of practical use, which they consider 
is
\be 
A_{D-N}^s=2\frac{R^s-R^D}{R^s+R^D},
\label{eq:asym}
\ee   
where $R^s$ is the counting rate in SuperKamiokande for the neutrino events 
with the energies of the recoil electrons higher than $5~MeV$ in the night 
time, when neutrino passes through the core ($s=c$) or through the mantle 
only ($s=m$) and $R^D$ is the counting rate in the day time. Therefore,
$A_{D-N}^s$ is the asymmetry related to the counting rate of neutrinos 
crossing the earth core or mantle only. The calculations are performed for 
the wide range of $\Delta m^2$ and $\sin^22\theta$, but the most interest 
should be given to the values corresponding to the small-angle and 
large-angle MSW solutions. For most interesting small-angle solution, 
the asymmetry is much stronger for the core case. The asymmetry here varies 
between $1\%$ and $20\%$, but in some cases it is less than $1\%$ and even
zero. One should remember that statistics of the core sample is low and 
long observations are needed to discover $1\%$ asymmetry.\\*[1mm]

The anomalous seasonal variation of neutrino flux is a signature of 
{\em vacuum oscillations}. The distance between the sun and earth changes 
with time as
\be
r(t)= r_0(1+\epsilon \cos 2\pi t/T),
\label{eq:geom}
\ee
where  $\epsilon = 0.0167$ is the ellipticity of the earth orbit and 
$r_0=1.5 \cdot 10^{13}~cm$ is the mean distance between the sun and earth.
Thus, the geometrical seasonal variation in neutrino fluxes is about $7\%$
between the maximum and minimum. 

The time variation of distance between the sun and earth, results in  
the time dependent conversion $\nu_e \to \nu_{\mu}$, as one can see from 
Eq.(\ref{eq:prob}), where $r$ now is a function of time. The time variation 
of solar neutrino flux was studied in the detail by Krastev and Petkov 
(1994) (for the recent calculations see Smirnov 1997). The strongest 
seasonal variation was found for $^7Be$ and pep-neutrinos. These variations 
much exceed the geometrical one, and they will be unmistakably tested by 
Borexino. The seasonal variations of boron neutrino flux due to vacuum 
oscillations is comparable with 
the geometrical variations. For some allowed regions 
($\Delta m^2, sin^22\theta$) they compensate the geometrical
variations, for others they increase them by factor of 2. The variation 
of the flux correlates with spectrum distortion (Smirnov 1997). 
SuperKamiokande can 
discover the flux variation caused by  "central" vacuum solution during 
5 years of observations (Y.Suzuki, private communication).\\*[1mm]

The signature of the RSFP-solution is time variation of solar neutrino flux 
with the period of variation of magnetic field in the sun. 

The magnetic 
field responsible for the RSFP is most probably the toroidal magnetic field 
at the bottom of convective zone. The magnetic activity of the sun  exhibits 
quasi-periodic time variations with the mean period $11~yr$. Taking into 
account changing of magnetic polarity, one can argue for $22~yr$ as a 
basic period for large-scale magnetic field. This periodicity 
is thought to be originated due to toroidal field, generated in so-called
{\em overshoot layer} by dynamo mechanism and located near the bottom of 
convective zone. Theoretically, magnetic field there can reach $100~kG$.
This field rises through convective zone to the surface of the sun.

The RSFP-effect for beryllium and boron neutrinos takes place in the 
convective zone, while for pp-neutrinos in the radiative zone, where magnetic 
field is known worse. 
As was mentioned before, the RSFP solution needs magnetic field in the 
convective zone in the range $25-130~kG$, which is allowed by the 
theoretical concept described above. 

The time variation of magnetic field results in the framework of the RSFP 
scenario 
in the variation of neutrino flux with the same period. The variation 
of neutrino flux depends on neutrino energy. This dependence one 
can see from energy dependence of the suppression factor. It is 
straightforward that variation is 
stronger for energies, where suppression is stronger. Therefore, 
pp-neutrinos should not vary with time, while the boron neutrinos 
and especially $^7Be$-neutrinos must expose time variation.

The straightforward analysis of neutrino signals in all 
four solar-neutrino detectors does not reveal statistically significant 
time variations. In particular, it is true for the Homestake data
(Lissia 1996a and 1996b). However, when correlations with surface solar phenomena 
such as surface magnetic field, solar spots, green line and solar wind are 
included in the analysis,
the 11 year time variations of neutrino signal in the Homestake data become 
statistically 
significan6t, at least up to 1990 (for a comprehensive review see 
Stanev 199a and 1996b). This situation reminds me the epic of Cyg X-3, 
from which the 
direct signal in high energy gamma-rays in all cases, except the Kiel detector, 
was not seen, while 
the correlation with $4.8~h$ periodical X-ray flux, gave the statistically 
significant signal in many detectors.

One of the objections against the time-variable signal in the Homestake 
detector is the absence of such dependence in the Kamiokande and 
gallium detectors. This objection actually is not correct. The time 
variation effect in the Kamiokande detector, in case of the RSFP, must be 
weaker than in the the Homestake detector. First, the Homestake detector 
has significant, $13\%$, contribution from strongly variable flux of 
$^7Be$-neutrinos. Secondly, $\nu_{\mu}$-neutrinos from oscillation of $\nu_e$ 
give contribution to the Kamiokande signal, but not to the 
Homestake signal (Akhmedov 1997). 

The time variation of the signal should 
exist for both gallium experiments (due to $28\%$ contribution of the 
beryllium-neutrino flux) and for SuperKamiokande, because even at 
highest energies the boron neutrino flux under the RSFP is suppressed by 
factor not smaller than 2. However, this dependence is expected to be 
weaker than in the Homestake detector.

The Borexino detector, observing directly beryllium neutrinos, will 
reliably test the abovementioned time variation. 
  
Another possible signature of the RSFP is production of 
$\bar{\nu}_e$-neutrinos.
They appear due to vacuum oscillation of $\bar{\nu}_{\mu}$ on the way from 
the sun to earth (see Akhmedov 1997 and the references therein). As one 
can see from Eq.(\ref{eq:prob}) the amplitude 
of this oscillation is proportional to $sin^22\theta$. For a typical 
RSFP solution this value is 0.1-0.2 and thus the $\bar{\nu}_e$-flux 
with the energy of beryllium neutrinos will be easily detectable in 
the Borexino.\\*[3mm]
\noindent
7. CONCLUSIONS\\*[3mm]
\noindent
The Solar Neutrino Problem (SNP) is a deficit of neutrino fluxes detected 
in all four solar-neutrino experiments, as compared with the SSMs 
predictions. Not related to the SSMs, there is a conflict  
between data of neutrino experiments, which within astrophysical solution
to the SNP (standard neutrino), results in unphysical zero beryllium 
neutrino flux.

This result was known for the last several years. The situation has changed 
 now when the predictions of the SSMs (sound speed and density 
radial profiles) are confirmed by the precise helioseismic measurements
for all radial distance at interest. The recent measurements of 
cross-section $^3He+^3\!\!He \to ^4\!\!He+2p$ at the energy of the 
solar Gamow peak in the LUNA experiment at Gran Sasso, confirmed further 
the predictions of solar-neutrino fluxes by the SSMs. It is difficult to 
doubt now that the SSMs give the adequate description of the sun 
interior. The conflict between the observed neutrino flux and the 
predictions of seismically confirmed SSMs can be considered as the 
strongest argument against the standard neutrino. 

The SSMs being sufficient for prediction of neutrino fluxes are unable
to describe the small irregularities in the observed radial profiles 
of density and sound speed. They imply some physical processes not 
included in the SSMs, which most probably result in the small 
mixing inside the sun not much relevant to the prediction of neutrino 
fluxes.  

30-40 years ago we thought that neutrinos with their tremendous penetrating 
power give us the only way to look inside the sun.  Now we see that 
helioseismology makes it more precisely, while the solar neutrino fluxes 
give us unique information about neutrino properties.

There are three elementary-particle solutions to the SNP: the MSW,
vacuum oscillations and the Resonant Spin-Flavor Precession (RSFP). In all 
three solutions the neutrino is not standard, i.e. it is not described by the 
SM of
EW interactions. It is massive, the lepton numbers are not conserved and 
the mass matrix is not diagonal in the flavor representation. 

The elementary particle solutions successfully describe the results of all 
four solar-neutrino experiments, because the suppression factors of 
neutrino fluxes are energy dependent (the beryllium neutrino flux is 
most strongly suppressed). It immediately results in the common 
signature of all three solutions: the measured boron-neutrino energy spectrum 
must be distorted as compared with the $^8B$-decay neutrino spectrum.
It looks most probable that SuperKamiokande and SNO will discover 
the distortion of the spectrum in the near future. The additional signature 
could be the anomalous neutral-current scattering (through neutron detection) 
in SNO, and measurement of $Be$-neutrino flux in Borexino.

 To distinguish between these three solutions is much more difficult task.

The signature of the MSW-solution is the day-night effect, which can be 
observed by SuperKamiokande. The value of 
this effect can vary between $1\%$ and $20\%$, being zero for some 
values of ($\Delta m^2,\sin^22\theta$). In all cases, the effect is 
strongest when neutrinos cross the earth core. It limits 
significantly statistics of observations.   

The signature of the vacuum oscillations is a seasonal variation of 
neutrino flux,  in addition to geometrical seasonal variation $1/r^2(t)$,
where $r(t)$ is a time dependent distance between the earth and sun.
These variations are very significant for beryllium neutrinos and can  be 
reliably tested by Borexino.

The signature of the RSFP-solution is the 11 year periodicity of neutrino 
flux with the largest amplitude for beryllium neutrinos. Another signature 
is a considerable fraction of $\bar{\nu}_e$ neutrinos at the energy of 
beryllium neutrinos due to vacuum oscillation 
$\bar{\nu}_{\mu} \to \bar{\nu}_e$ on the way from the sun to earth. Both 
signatures can be reliably detected by Borexino.

The next generation detectors, ICARUS and HELLAZ, will provide us with 
much more detailed information about energies of neutrinos and arrival 
direction.\\*[3mm]
\noindent
ACKNOWLEDGEMENTS\\*[3mm]
\noindent
I am greatly impressed by excellent organization of 25th ICRC in Durban 
and by tremendous effiiciency of all members of the Organizing Committee.
Special thanks are to Harm Moraal for his patience to settle my personal 
problems. 

I have great pleasure to thank my collaborator of many-years Gianni Fiorentini
for many helpful discussions and advices. Many thanks are to all other 
people with whom I am working on the solar-neutrino subject and to whom 
I am indebted for better understanding of these problems, namely, to 
V.Castellani, S. Degl'Innocenti, W. Dziembowski, M.Lissia and B.Ricci.
I am grateful to J.Christensen-Dalsgaard for valuable remarks and for 
sending me the figures from his works. 
And finally, I am greatly indebted to 
N.Hata, M.Junker and P.Krastev who have kindly provided 
me with the figures from their works.\\*[3mm]
\noindent

REFERENCES\\*[1mm]

\noindent
Abdurashitov  J.N. et al (SAGE collaboration) 1996, 
Phys. Rev. Lett. {\bf 77}, 4708.\\

\noindent
Arpesella C. et al (LUNA collaboration) 1996, Phys. Lett. {\bf B 389}, 452.
\\

\noindent
Akhmedov E.Kh. 1988, Phys. Lett. {\bf B 213}, 64.\\

\noindent
Akhmedov E.Kh., Lanza A. and Petcov S.T. 1993, Phys. Lett. {\bf B 303},85.
\\

\noindent
Akhmedov E.Kh. 1997,  Invited talk given at 4th Int. Solar Neutrino Conf.,
Heidelberg, April 8-11,1997, hep-ph/97055451.\\

\noindent
Bahcall J.N. 1989, Neutrino Astrophysics, Cambridge Univ. Press, 
Cambridge.\\

\noindent
Bahcall J.N. and Bethe H.A. 1990, Phys. Rev. Lett. {\bf 65}, 2233.\\

\noindent
Bahcall J.N. 1993, Phys. Rev. Lett. {\bf 71}, 2369.\\

\noindent
Bahcall J.N. 1994, Phys. Lett. {\bf B 338}, 276.\\

\noindent
Bahcall J.N. and Pinsonneault 1995, Review of Modern Physics, {\bf 67},781.\\

\noindent
Bahcall J.N., Pinsonneault M.H., Basu S., and Christensen-Dalsgaard J. 1997
, Phys. Rev. Lett.  {\bf 78}, 171.\\

\noindent
Bahcall J.N. and Krastev P. 1997, Preprint IASSNS-AST 97/31 to be published 
in Phys. Rev.\\ 

\noindent
Barbieri R. and Fiorentini G. 1998, Nucl. Phys. {\bf B 304}, 909.\\

\noindent
Bellini G. et al 1996, Nucl. Phys. B (Proc. Suppl.), {\bf 48}, 363;
Arpesella C. et al, "Borexino at Gran Sasso"(Proposal),
ed. by G.Bellini and R.Raghavan, INFN, Univ.of Milan, 1991;
Bellotti E., Nucl.Phys. B (Proc. Suppl){\bf 38}, 90, (1995).\\

\noindent
Berezinsky V. 1994, Comm. Nucl. Part. Phys. {\bf 21}, 249\\

\noindent
Berezinsky V., Fiorentini G. and Lissia M. 1994, Phys. Lett. {\bf B341},
38.\\

\noindent
Berezinsky V. 1995, Nuovo Cim. {\bf 18 C}, 671.\\

\noindent
Berezinsky V., Fiorentini G. and Lissia M. 1996, Phys. Lett. {\bf B365},
185.\\

\noindent
Bludman S., Hata N., Kennedy D. and Langacker P. 1993, Phys. Rev. 
{\bf D 47}, 2220.\\

\noindent
Bowles T.J. and Gavrin V.N. 1993, Annu. Rev. Nucl. Part. Sci. {\bf 43},
117.\\

\noindent
Brown L.S. and Sawyer R.F. 1997, astro-ph/9704299.\\

\noindent
Castellani V., Degl'Innocenti S., Fiorentini G. 1993, 
Astron. Astrophys. {\bf 271}, 601.\\

\noindent
Castellani V., Degl'Innocenti S., Fiorentini G., Lissia M., Ricci B. 1997,
Phys. Rep. {\bf 281}, 309.\\

\noindent
Christensen-Dalsgaard J. 1996a, Nucl. Phys. B (Proc. Suppl.) {\bf 48},
325.\\

\noindent
Christensen-Dalsgaard J. 1996b, Proc. of 18th Texas Symposium on Relativistic 
Astrophysics (Chicago) to be published.\\

\noindent
Christensen-Dalsgaard J.et al  1996c, Science {\bf 272}, 1286.\\ 

\noindent
Cleveland B.T et al (Homestake) 1995,  Nuclear Phys.B (Proc.Suppl.)
{\bf 39}, 47.\\

\noindent
Conner Z. 1997, Highlight talk at 25th ICRC.\\

\noindent
Cumming A. and Haxton W.C. 1996, Phys. Rev. Lett. {\bf 77}, 4286.\\

\noindent
Degl'Innocenti S.,Fiorentini G. and Lissia M. 1995, {\bf 43}, 66.\\

\noindent
Degl'Innocenti S., Dziembowski W.A., Fiorentini G. and Ricci B. 1997a,
Astrop. Phys. {\bf 7}, 77.\\

\noindent
Degl'Innocenti S., Ciaco F. and Ricci B. 1997b, Astr. Astroph. Suppl. Ser.
{\bf 123}, 1.\\

\noindent
Dilke F.W.W. and Gough D.O. 1972, Nature {\bf 240}, 262.\\

\noindent
Dziembowski W.A. 1996, Bull. Astron. Soc. India, {\bf 24}, 133.\\

\noindent
Gough D.O. and Thompson M.J. 1991, In: Solar Interior and Atmosphere 
(ed.s Cox A.N. et al), University of Arizona Press, 519.\\

\noindent
Gruzinov A.V. and Bahcall J.N. 1997, astro-ph 9702065.\\

\noindent
Hampel W. et al (GALLEX collaboration), Phys.Lett. B 
388, 384, (1996).\\

\noindent
Hata N., Bludman S. and Langacker P. 1994, Phys.Rev.{\bf D 49}, 3622.\\

\noindent
Hata N. and Langacker P.  1997, hep-ph/9705339, to be published in PR D.\\

\noindent
Huber M. 1997, Invited lecture at 25th ICRC.\\

\noindent
ICARUS Collaboration 1994, Proposal v.1 and v.2, Laboratori Nazionali del 
Gran Sasso 94/99.\\

\noindent
Kirsten T. 1995, Proc. of  17th TEXAS Symposium on 
Relativistic Astrophysics (eds. H.Boeringer, G.E.Morfil and J.E.Truemper), 
Ann. N.Y. Academy of Sciences , 1.\\

\noindent
Krastev P.I. and Petcov S.T. 1995, Nucl. Phys. {\bf B 449}, 605.\\

\noindent
Krastev P.I. and Petcov S.T. 1996, Phys. Rev. {\bf D53}, 1665.\\

\noindent
Krastev P.I. and Smirnov A.Yu. 1994, Phys.Lett. {\bf B 338}, 282.\\

\noindent
Kwong W. and Rosen S.P. 1994, Phys. Rev. Lett. {\bf 73}, 369.\\

\noindent
Lim C.S. and Marciano W.J. 1988, Phys. Rev. {\bf D 37}, 1368.\\

\noindent
Lim C.S. and Nunokawa H. 1995, Astrop. Phys. {\bf 4}, 63.\\

\noindent
Lissia M. 1996a, Proc. 4th Int. Topical Workshop "New Trends in Solar 
Neutrino Physics" (eds. V.Berezinsky and G. Fiorentini), Gran Sasso,
Italy, p.129.\\

\noindent
Lissia M. 1986b, Proc. of 17th Texas Symposium on Relativistic Astrophysics
, Chicago, to be published.\\

\noindent
Maris M. and Petcov S.T. 1997, hep-ph/9705392.\\

\noindent
Mikheev S.P. and Smirnov A.Yu. 1986a, Nuovo Cim. {\bf 9 C}, 17.\\

\noindent
Mikheev S.P. and Smirnov A.Yu. 1986b, Proc. of the 6th Moriond Workshop 
(ed. J. Tran Thanh), p. 355.\\ 

\noindent
Mikheev S.P. and Smirnov A.Yu. 1986c, Sov. Phys. Uspekhi {\bf 30}, 759.\\

\noindent
Nunokawa H., Rossi A., Semikoz V.B. and Valle J.W.F. 1996, Nucl. Phys.
{\bf B 472}, 495.\\

\noindent
Pontecorvo B. 1957, ZhETP {\bf 33}, 549.\\

\noindent
Raffelt G.G. 1990, Phys. Rev. Lett. {\bf 64}, 2856.\\

\noindent
Raghavan R.S. 1997, Phys. Rev. Lett. {\bf 78}, 3618.\\

\noindent
Ricci B., Berezinsky V., Degl'Innocenti S., Dziembowski W. and 
Fiorentini G. 1997, Phys. Lett. {\bf B 407}, 155.\\

\noindent
Schechter J. and Valle J.W.F. 1981, Phys. Rev. {\bf D 24}, 1883.\\

\noindent
Smirnov A.Yu. 1997, talk at TAUP, September 1997.\\

\noindent
Stanev T. 1986a, Proc. 4th Int. Topical Workshop "New Trends in Solar 
Neutrino Physics" (eds. V.Berezinsky and G. Fiorentini), Gran Sasso,
Italy, p.141.\\

\noindent
Stanev T. 1986b, Proc. of 17th Texas Symposium on Relativistic Astrophysics
, Chicago, to be published.\\

\noindent
Suzuki Y., Rapporteur talk at 25th ICRC (Durban) 1997.\\

\noindent
Tsytovich V.N., Bingham R., De Angelis U., Forlani A. and Occorsio M.R.
1996, Astroparticle Physics {\bf 5}, 197.\\

\noindent
Turck-Chieze S., D\"appen W., Fossat E., Provost J., 
Schatzman E. and Vignaud D. 1993 , Phys. Rep. {\bf 230}, 57, (1993).\\

\noindent
Voloshin M.B., Vysotsky M.I. and Okun L.B. 1986, Sov. J. Nucl. Phys.
{\bf 44}, 845.\\

\noindent
Wolfenstein L. 1978, Phys.Rev. {\bf D17}, 2369.\\

\noindent
Ypsilantis T. 1982, Proc. 4th Int. Workshop "Neutrino Telescopes 
(ed. M. Baldo-Ceolin).\\

\end{document}